\documentclass[10pt]{article}
\usepackage[OE]{express}
\usepackage{amsmath}
\usepackage{verbatim}

\begin{document}
\title{Method for Computationally Efficient Design of Dielectric Laser Accelerator Structures}

\author{Tyler Hughes,\authormark{1} Georgios Veronis,\authormark{2} Kent~P.~Wootton,\authormark{3} R.~Joel~England,\authormark{3} and Shanhui~Fan\authormark{4,*}}

\address{\authormark{1}Department of Applied Physics, Stanford University, 348 Via Pueblo, Stanford, CA, 94305, USA\\
\authormark{2}School of Electrical Engineering and Computer Science and
Center for Computation and Technology (CCT), Louisiana State University, Baton Rouge, LA, 70803, USA\\
\authormark{3}SLAC National Accelerator Laboratory, 2575 Sand Hill Rd, Menlo Park, CA, 94025, USA\\
\authormark{4}Department of Electrical Engineering, Stanford University, 350 Serra Mall, Stanford, CA 94305, USA}

\email{\authormark{*}shanhui@stanford.edu}

\begin{abstract}
Dielectric microstructures have generated much interest in recent years as a means of accelerating charged particles when powered by solid state lasers.  The acceleration gradient (or particle energy gain per unit length) is an important figure of merit. To design structures with high acceleration gradients, we explore the adjoint variable method, a highly efficient technique used to compute the sensitivity of an objective with respect to a large number of parameters.  With this formalism, the sensitivity of the acceleration gradient of a dielectric structure with respect to its entire spatial permittivity distribution is calculated by the use of only two full-field electromagnetic simulations, the original and `adjoint'.  The adjoint simulation corresponds physically to the reciprocal situation of a point charge moving through the accelerator gap and radiating.  Using this formalism, we perform numerical optimizations aimed at maximizing acceleration gradients, which generate fabricable structures of greatly improved performance in comparison to previously examined geometries. \\
\end{abstract}

%\ocis{(230.0230)   Optical devices; (050.2770)   Gratings; (050.6624)   Subwavelength structures
%.} % REPLACE WITH CORRECT OCIS CODES FOR YOUR ARTICLE, MINIMUM OF TWO; Avoid using the OCIS codes for “General” or “General science” whenever possible.
%For a complete list of OCIS codes, visit: https://www.osapublishing.org/oe/submit/ocis/

%%%%%%%%%%%%%%%%%%%%%%% References %%%%%%%%%%%%%%%%%%%%%%%%%
%\begin{thebibliography}{99}
%\bibliographystyle{osajnl}
%\bibliography{Refs2}
%\end{thebibliography}
%\begin{comment}

%\end{comment}
%%%%%%%%%%%%%%%%%%%%%%%%%%  body  %%%%%%%%%%%%%%%%%%%%%%%%%%

\section{Introduction}
\label{sec:introduction}
% 1. Context of the study
Dielectric laser accelerators (DLAs) are periodic dielectric structures that, when illuminated by laser light, create a near-field that may accelerate electrically charged particles such as electrons \cite{england2014dielectric}. A principal figure of merit for these DLA structures is the acceleration gradient, which signifies the amount of energy gain per unit length achieved by a particle that is phased correctly with the driving field. DLAs may sustain acceleration gradients on the order of ${\sim}\textrm{GV}\,\textrm{m}^{-1}$ when operating using the high peak electric fields supplied by ultrafast (femtosecond) lasers. These acceleration gradients are several orders of magnitude higher than conventional particle accelerators.  As a result, the development of DLA can lead to compact particle accelerators that enable new applications. 

% 2. Statement of the problem
In previous works, candidate DLA geometries were optimized for maximum acceleration gradient by scanning through parameters of a specified structure geometry \cite{plettner2006proposed, peralta2013demonstration, mcneur2016elements, leedle2015dielectric, chang2014silicon, breuer2014dielectric, breuer2014dielectric2, kozak2016dielectric}. However, this strategy has limited potential to produce higher acceleration gradient structures because it only searches a small portion of the total design space.

% 3. Aim and scope
In this paper, we derive an analytical form for the sensitivity of the acceleration gradient of a DLA structure with respect to its permittivity distribution using the adjoint-variable method (AVM).  We may calculate this by use of only two full-field simulations. The first corresponds to the typical accelerator setup, where the structure is illuminated with externally incident laser light. The second corresponds to the inverse process, where the same physical structure is simulated but now with a charged particle traversing the structure as the source.  Thus, this formalism explicitly makes use of the reciprocal relationship between accelerators and radiators \cite{palmer_1995, joshi_2012}. We use this sensitivity information to perform optimizations, which generate DLA structures of much higher gradients than previously explored geometries.

% 4. Significance of the problem
This work is the first application of the AVM technique to the design of DLA structures and gives examples of fabricable structures that may  improve the energy gain achievable with current DLA technology.  In addition, the optimized structures give insight into general design principles for DLAs, meaning that one may use the principle findings of this paper to design DLAs without having to run optimizations directly.  As an example, it was found that high gradient structures often include dielectric mirrors surrounding the particle gap, leading to higher field enhancement.

% 5. Overview of paper
This paper is organized as follows:  We first outline the status of DLAs and basic design requirements in section 2.  We introduce AVM in section 3, where we derive the sensitivity of the acceleration gradient of a DLA with respect to its permittivity distribution.  In section 4, we show  that the `adjoint' solution corresponds to that of a radiating charge.  In section 5, we describe and demonstrate algorithms for using the sensitivity information to design DLA structures numerically.

\section{A Brief Review of Dielectric Laser Accelerators}
\label{sec:dla}
DLAs take advantage of the fact that dielectric materials have high damage thresholds at short pulse durations and infrared wavelengths \cite{england2014dielectric, mcneur2016elements, soong2012laser} when compared to metal surfaces at microwave frequencies.  This allows DLAs to sustain peak electromagnetic fields, and therefore acceleration gradients, that are 1 to 2 orders of magnitude higher than those found in conventional radio frequency (RF) accelerators.  Experimental demonstrations of these acceleration gradients have been made practical in recent years by the availability of robust nanofabrication techniques combined with modern solid state laser systems \cite{dawson2008analysis}.  By providing the potential for generating relativistic electron beams in relatively short length scales, DLA technology is projected to have numerous applications where tabletop accelerators may be useful, including medical imaging, radiation therapy, and X-ray generation \cite{plettner2008microstructure,england2014dielectric}.  To achieve high energy gain in a compact size, it is of principle interest to design structures that may produce the largest acceleration gradients possible without exceeding their respective damage thresholds.

\begin{figure}[htb!]
\centering\includegraphics[width=\textwidth]{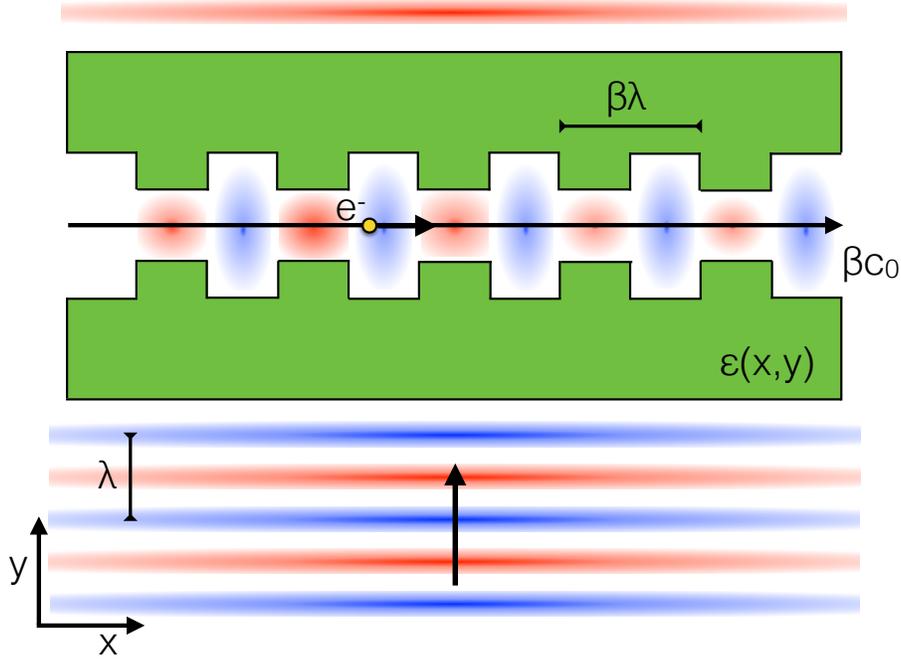}
\caption{Diagram outlining the system setup for side-coupled DLA with an arbitrary dielectric structure $\epsilon(x,y)$ (green).  A charged particle moves through the vacuum gap with speed $\beta c_0$.  The periodicity is set at $\beta \lambda$ where $\lambda$ is the central wavelength of the laser pulse.}
\label{fig:system}
\end{figure}

Several recently demonstrated candidate DLA structures consist of a planar dielectric structure that is periodic along the particle axis with either an semi-open geometry or a narrow (micron to sub-micron) vacuum gap in which the particles travel \cite{plettner2006proposed, peralta2013demonstration, mcneur2016elements, leedle2015dielectric, chang2014silicon, breuer2014dielectric, breuer2014dielectric2, kozak2016dielectric}.  These structures are then side-illuminated by laser pulses. Fig.~\ref{fig:system} shows a schematic of the setup, with a laser pulse incident from the bottom.

The laser field may also be treated with a pulse front tilt \cite{hebling1996derivation, akturk2004pulse} to enable group velocity matching over a distance greater than the laser's pulse length.  For acceleration to occur, the dielectric structure must be designed such that the particle feels an electric field that is largely parallel to its trajectory over many optical periods. In the following calculations, the geometry of the dielectric structure is represented by a spatially varying dielectric constant $\epsilon(x,y)$.  We assume invariance in one coordinate ($\hat{z}$) in keeping with the planar symmetry of most current designs.  However the methodology we present can be extended to include a third dimension.  In addition, our work approximates the incident laser pulse as a monochromatic plane wave at the central frequency, which is a valid approximation as long as the pulse duration is large compared to the optical period.

\section{Adjoint Variable Method}
\label{sec:examples}

In a general DLA system, we may define the acceleration gradient `$G$' over a time period `$T$' mathematically as follows:

\begin{equation}
G = \frac{1}{T}\int_0^{T}{ E_{||}(\vec{r}(t),t)\ dt}
\label{eq:Gintro}
\ ,
\end{equation} 
where $\vec{r}(t)$ is the position of the electron and $E_{||}$ signifies the (real) electric field component parallel to the electron trajectory at a given time.

To maximize this quantity, we employ AVM \cite{georgieva2002feasible,bakr2003adjoint}, which is a technique common to a wide range of fields, including seismology \cite{plessix2006review}, aircraft design \cite{giles2000introduction}, and, recently, photonic device design \cite{veronis2004method, lalau2013adjoint, piggott2015inverse}.  Many engineering systems can be described by a linear system of equations $A(\gamma) z = b$, where $\gamma$ is a set of parameters describing the system. For a given set of parameters $\gamma$, solving this equation results in the solution `$z$', from which an objective $J=J(z)$, which is a function of the solution, can be constructed. The optimization of the engineering system corresponds to maximizing or minimizing $J$ with respect to the parameters $\gamma$. For this purpose, AVM allows one to calculate the gradient of the objective function $\nabla_\gamma J$ for an arbitrary number of parameters $\gamma_i$ with the only added computational cost of solving one additional linear system $\hat{A}^T \bar{z} = -\frac{dJ}{dz}^T$, which is often called the `adjoint' problem.  For a more comprehensive overview of the method, we refer the reader to \cite{georgieva2002feasible}.

Here we provide the derivation of AVM specifically for the optimization of the accelerator structures. Since the structure is invariant in the $\hat{z}$ direction, we work in two dimensions, examining only the $H_z$, $E_x$ and $E_y$ field components.  For an approximately monochromatic input laser source with angular frequency $\omega$, the electric fields are, in general, of the form

\begin{equation}
\vec{E}(\vec{r},t) = \textrm{Re}\left\{\vec{E}(\vec{r})\ \textrm{exp}(i\omega t)\right\},
\end{equation}
where now $\vec{E}$ is complex.

Let us assume the particle we wish to accelerate is moving on the line $y=0$ with velocity $\vec{v} = \beta c_0 \hat{x}$, where $c_0$ is the speed of light in vacuum and $\beta \leq 1$.  The $x$ position of the particle as a function of time is given by $x(t) = x_0 + \beta c_0 t$, where $x_0$ represents an arbitrary choice of initial starting position.  For normal incidence of the laser (laser propagating in the $+\hat{y}$ direction), phase velocity matching between the particle and the electromagnetic fields is established by introducing a spatial periodicity in our structure of period $\beta \lambda$ along $\hat{x}$ , where $\lambda$ is the laser wavelength.  In the limit of an infinitely long structure (or equivalently, $T \to \infty $) we may rewrite our expression for the gradient in Eq. (\ref{eq:Gintro}) as an integral over one spatial period, given by

\begin{equation}
G = \frac{1}{\beta\lambda}\textrm{Re}\left\{ \textrm{exp}(-i\phi_0) \int_0^{\beta\lambda}{dx \ }E_x(x,0)\ \textrm{exp}\Big( i\frac{2\pi}{\beta\lambda}x \Big) \right\}.
\end{equation}
Here the quantity $\phi_0 = \frac{2\pi x_0}{\beta\lambda}$ is representative of the phase of the particle as it enters the spatial period.  In further calculations, we set $\phi_0 = 0$, only examining the acceleration gradients experienced by particles entering the accelerator with this specific phase.  Since we have arbitrarily control over our input laser phase, this does not impose any constraint on the acceleration gradient attainable.

To simplify the following derivations, we define the following inner product operation involving the integral over two vector quantities $\vec{a}$ and $\vec{b}$ over a single period volume $V'$

\begin{equation}
\langle \vec{a},\vec{b} \rangle = \langle \vec{b},\vec{a} \rangle = \int_{V'} dv \  \left(\vec{a}\cdot\vec{b}\right) = \int_0^{\beta\lambda}dx\int_{-\infty}^\infty dy \   \left(\vec{a}\cdot\vec{b}\right).
\end{equation} 
With this definition, we then have the gradient

\begin{equation}
G = \textrm{Re}\{ \langle \vec{E}, \vec{\eta} \rangle \},
\label{eq:G}
\end{equation} 
where

\begin{equation}
\vec{\eta} = \vec{\eta}(x,y) = \frac{1}{\beta\lambda}\textrm{exp}\Big(i\frac{2\pi}{\beta\lambda}x\Big)\delta(y)\hat{x}.
\end{equation}

Now, we wish to examine the sensitivity of $G$ with respect to an arbitrary parameter, $\gamma$, which may represent a shifting of material boundary, changing of dielectric constant at a point, or any other change to the system.  Differentiating Eq. (\ref{eq:G}) gives

\begin{equation}
\frac{dG}{d\gamma} = \textrm{Re}\left\{ \left\langle \frac{d\vec{E}}{d\gamma}, \vec{\eta} \right\rangle \right\}
\label{eq:dGdgamma}.
\end{equation} 
Here we have made use of the fact that $\vec{\eta}$ does not depend on $\gamma$.

From Maxwell's equations in the frequency domain, we may express our electromagnetic problem in terms of a linear operator $\hat{A}$ as

\begin{equation}
\nabla \times \nabla \times\vec{E}(\vec{r})\ -\ k_0^2\ \epsilon_r(\vec{r})\ \vec{E}(\vec{r}) \equiv \hat{A}\vec{E}(\vec{r}) = -i\mu_0\omega\vec{J}(\vec{r})
\label{eq:FDFD_analytical}.
\end{equation}
Here, $k_0 = \omega/c_0$, $\epsilon_r$ is the relative permittivity, $\vec{J}$ represents a current density source, and a non-magnetic material is assumed ($\mu = \mu_0$).  Differentiating Eq. (\ref{eq:FDFD_analytical}) with respect to $\gamma$, and assuming that the current source ($\vec{J}$) does not depend on $\gamma$, we see that

\begin{equation}
\frac{d\vec{E}}{d\gamma} = -\hat{A}^{-1}\frac{d\hat{A}}{d\gamma}\vec{E}
\label{eq:dEdgamma}.
\end{equation}
$\hat{A}$ is self-adjoint under our inner product, $\langle \hat{A}\vec{a}, \vec{b} \rangle = \langle \vec{a}, \hat{A}\vec{b} \rangle$, and the same is true for $\hat{A}^{-1}$ and $\frac{d\hat{A}}{d\gamma}$.  Using these facts and combining Eq. (\ref{eq:dGdgamma}) with Eq. (\ref{eq:dEdgamma}), we find that

\begin{equation}
\frac{dG}{d\gamma} =  \textrm{Re} \left\{ \left\langle -\hat{A}^{-1}\frac{d\hat{A}}{d\gamma}\vec{E}\ , \vec{\eta}  \right\rangle \right\} = \textrm{Re}\left\{ \left\langle \vec{E}\ , -\frac{d\hat{A}}{d\gamma}\hat{A}^{-1}\vec{\eta} \right\rangle \right\}
\label{eq:dGdgamma_2}.
\end{equation}

Thus, if we define a second simulation with a source of $-\vec{\eta}$ and fields $\vec{E}_{aj}$, 

\begin{equation}
\hat{A}\vec{E}_{aj} = -i\mu_0\omega\vec{J}_{aj} = -\vec{\eta},
\label{eq:adjoint_source}
\end{equation}
then the field solution, $\vec{E}_{aj} = -\hat{A}^{-1}\vec{\eta}$, can be easily identified in Eq. (\ref{eq:dGdgamma_2}).  The sensitivity of the acceleration gradient can finally be expressed as

\begin{equation}
\frac{dG}{d\gamma} = \textrm{Re} \left\{ \left\langle \vec{E}\ , \frac{d\hat{A}}{d\gamma} \vec{E}_{aj} \right\rangle \right\}.
\label{eq:final_form_DGdgamma}
\end{equation}

The only quantity in this expression that depends on the parameter $\gamma$ is $\frac{d\hat{A}}{d\gamma}$.  As we will soon discuss, this quantity will generally be trivial to compute.  On the other hand, the full field calculations of $\vec{E}$ and $\vec{E}_{aj}$ are computationally expensive, but may be computed once and used for an arbitrarily large set of parameters $\gamma_i$. This gives the AVM approach a significant scaling advantage with respect to traditional direct sensitivity methods, which require a separate full-field calculation for each parameter being investigated.  It is this fact that we leverage with AVM to do efficient optimizations over a large design space.

To confirm that this derivation matches the results obtained by direct sensitivity analysis, we examine a simple accelerator geometry composed of two opposing dielectric squares each of relative permittivity $\epsilon$.  We take a single $\gamma$ parameter to be the relative permittivity of the entire square region.  Because we only change the region inside the dielectric square, we may identify the $\frac{d\hat{A}}{d\gamma}$ operator for this parameter by examining Eq.~(\ref{eq:FDFD_analytical}), giving

\begin{equation}
    \frac{d\hat{A}}{d\epsilon}(\vec{r}) = \Bigg\{
    \begin{array}{ll}
      -k_0^2 &\text{if } \vec{r} \text{ in square} \\
      0 &\text{otherwise}
      \end{array}.
    \label{eq:dAdepsilon}
\end{equation}

\begin{figure}[htb!]
\centering\includegraphics[width=12cm]{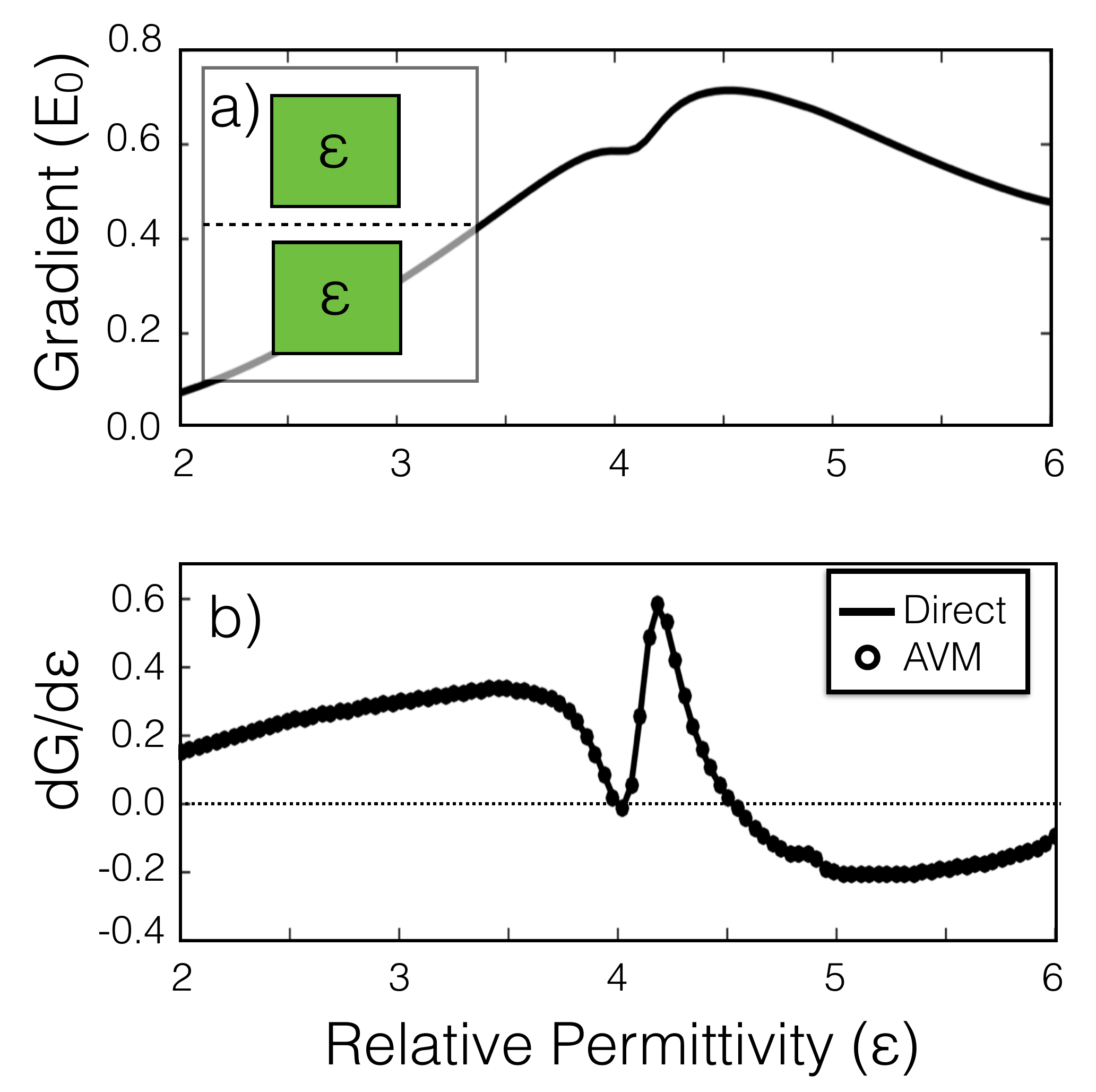}
\caption{Demonstration of AVM in calculating sensitivities.  (a) The acceleration gradient ($G$) of a square accelerator structure (inset) as a function of the square's relative permittivity.  The particle traverses along the dotted line and a plane wave is incident from the bottom of the structure.  (b) The sensitivity $\frac{dG}{d\epsilon}$ of the gradient with respect to changing the square relative permittivity for direct central difference (solid line) $\frac{dG}{d\epsilon} = \frac{G(\epsilon+\Delta\epsilon)-G(\epsilon-\Delta\epsilon)}{2\Delta\epsilon}$ and using AVM (circles).  The two calculations agree with excellent precision.  The dotted line at $\frac{dG}{d\epsilon}=0$, corresponds to local minima and maxima of $G(\epsilon)$ above.}
\label{fig:AVM-test}
\end{figure}

Thus, given the form of the acceleration gradient sensitivity given in Eq. (\ref{eq:final_form_DGdgamma}), combined with Eq. (\ref{eq:dAdepsilon}), the change in acceleration gradient with respect to changing the entire square permittivity is simply given by the integral of the two field solutions over the square region, labeled `$sq$'

\begin{equation}
\frac{dG}{d\epsilon_{sq}} = -k_0^2 \ \textrm{Re} \left\{ \int_{sq}{d^2\vec{r}}\  \vec{E}(\vec{r}) \cdot \vec{E}_{aj}(\vec{r}) \right\}.
\end{equation}
In Fig. \ref{fig:AVM-test} we compare this result with the direct sensitivity calculation where the system is manually changed and simulated again.  The two methods agree with excellent precision.

Extending this example to the general case of perturbing the permittivity at an arbitrary position $\vec{r}'$, we see that

\begin{align}
\frac{dG}{d\epsilon}(\vec{r}') &= -k_0^2 \ \textrm{Re}  \left\{  \int{d^2\vec{r}}\  \vec{E}(\vec{r}) \cdot \vec{E}_{aj}(\vec{r})\  \delta(\vec{r}-\vec{r}') \right\} \\
&= -k_0^2 \ \textrm{Re}  \left\{ \vec{E}(\vec{r}') \cdot \vec{E}_{aj}(\vec{r}') \right\}.
\end{align}

\section{Reciprocity}

With the AVM form derived, we now wish to re-examine the adjoint source term from Eq. (\ref{eq:adjoint_source}) in another interpretation.  Let us now consider the fields radiated by a point particle of charge $q$ flowing through our domain at $y = 0$ with velocity $\vec{v} = \beta c_0 \hat{x}$.  In the time domain, we can represent the current density of this particle as 

\begin{equation}
\vec{J}_{rad}(x,y;t) = q \beta c_0 \delta(x-x_0-c_0\beta t)\delta(y)\hat{x}.
\label{eq:point-current-setup}
\end{equation}

We may take the Fourier transform of $\vec{J}_{rad}$ with respect to time to examine the current density in the frequency domain, giving

\begin{align}
\vec{J}_{rad}(x,y;\omega) &= q \beta c_0 \delta(y)\hat{x}\int _{-\infty}^{\infty}{dt\ }\textrm{exp}(-i\omega t)\delta(x-x_0-c_0\beta t)\\
&= q\ \textrm{exp}\Big(i\frac{\omega\ (x-x_0)}{c_0 \beta}\Big)\delta(y)\hat{x}\\
&= q\ \textrm{exp}\Big(i\frac{2\pi}{ \beta\lambda}x\Big)\ \textrm{exp}(-i\phi_0)\delta(y)\hat{x}.
\label{eq:point-current}
\end{align}
Comparing with the source of our adjoint problem, $\vec{J}_{aj} = \frac{-i}{\omega\mu_0}\vec{\eta}$, we can see that 

\begin{equation}
\vec{J}_{aj} = \frac{-i\ \textrm{exp}(i\phi_0)}{2\pi q\beta c_0\mu_0}\vec{J}_{rad}.
\label{eq:J-eta}
\end{equation}

  \hfill \break

This finding shows that the adjoint field solution ($\vec{E}_{aj}$) corresponds (up to a complex constant) to the field radiating from a test particle flowing through the accelerator structure.  To put this another way, in order to calculate the acceleration gradient sensitivity with AVM, we must simulate the same structure operating both as an accelerator ($\hat{A}\vec{E}=-i\omega\mu_0\vec{J}_{acc}$) and as a radiator ($\hat{A}\vec{E}_{aj}=-i\omega\mu_0\vec{J}_{aj}$). 

It is understood that one way to create acceleration is to run a radiative process in reverse.  Indeed, this is the working principle behind accelerator schemes such as inverse free electron lasers \cite{musumeci2005high, courant1985high}, inverse Cherenkov accelerators \cite{kimura1995laser,fontana1983high}, and inverse Smith-Purcell accelerators \cite{bae1992experimental, mizuno1987experimental}.  Here, we see that this relationship can be expressed in an elegant fashion using AVM.

\section{Applications}

\subsection{Finite-Difference Frequency-Domain Modeling}

Now that we have shown how to use AVM to compute the sensitivity of the acceleration gradient with respect to the permittivity distribution, we will show practical applications of these results.  First, for computational modeling, the problem must be transitioned from a continuous space to a discrete space.  Here we make the transition using a finite-difference frequency-domain (FDFD) formalism \cite{shin2012choice, taflove2000computational}.  The electromagnetic fields now exist on a Yee lattice and the linear operator $\hat{A}$ becomes a sparse, complex symmetric matrix, $A$, relating the vector of electric field components, $\mathbf{e}$, to the input current source components $\mathbf{b}$ as

\begin{equation}
A\mathbf{e} = \mathbf{b}.
\label{eq:basic_FDFD}
\end{equation}

To solve for the field components, this system must be solved numerically for $\mathbf{e}$.  In two-dimensions, this is usually done directly by use of ``lower-upper'' (LU) decomposition methods for sparse matrices.  Only the right hand side of Eq. (\ref{eq:basic_FDFD}) is different between the original and adjoint simulations.  Therefore after the $A$ matrix is factored to solve the original simulation, its factored form may be saved and reused for the adjoint calculation, which cuts the total computational running time roughly in half.  

Written in terms of this discrete system, the acceleration gradient is

\begin{equation}
G = \rm{Re} \{\mathbf{e}^T \boldsymbol{\eta}\},
\end{equation}
where $\boldsymbol{\eta}$ is now a discretized version of $\vec{\eta}$.  Similarly, the sensitivity of the gradient with respect to changing the permittivity at pixel `$i$' is given by

\begin{equation}
\frac{dG}{d\epsilon_i} = -k_0^2\ \rm{Re} \{\mathbf{e}_{i}\bar{\mathbf{e}}_i\},
\end{equation}
where, as before, $\bar{\mathbf{e}}$ is the solution of the adjoint problem

\begin{equation}
A\bar{\mathbf{e}} = -\boldsymbol{\eta}.
\end{equation}

For all simulations, we use an FDFD program developed specifically for this work, although a commercial package would also be sufficient.  We have chosen a grid spacing that corresponds to 200 grid points per free space wavelength in each dimension.  Perfectly matched layers are implemented as absorbing regions on the edges parallel to the electron trajectory, with periodic boundary conditions employed on boundaries perpendicular to the electron trajectory.  A total-field scattered-field \cite{taflove2000computational} formalism is used to create a perfect plane wave input for the acceleration mode.

\subsection{Gradient Maximization}

Since we now have a highly efficient method to calculate $\frac{dG}{d\epsilon_{i}}$, we proceed to use this information to maximize the acceleration gradient with respect to the permittivity distribution.  We use an iterative algorithm based on batch gradient ascent \cite{avriel2003nonlinear}.  During each iteration, we first calculate $\frac{dG}{d\epsilon_{i}}$ for all pixels `$i$' within some specified design region.  Then, we update each $\epsilon_i$ grid as follows

\begin{equation}
\epsilon_{i} := \epsilon_{i} + \alpha\frac{dG}{d\epsilon_{i}}.
\end{equation}
% * <england@slac.stanford.edu> 2017-04-10T23:24:20.627Z:
% 
% For the comment on similarity to recently proposed structures, it would be good to put some kind of citation here.  Perhaps a talk or proceedings can be referenced?  If nothing like that exists, then it could just be cited as a "private communication."
% 
% ^.
Here, $\alpha$ is a step parameter that we can tune.  We need $\alpha$ to be small enough to find local maxima, but large enough to have the optimization run in reasonable amount of time. This process is then iterated until convergence on $G$.  During the course of optimization, the permittivity distribution is considered as a continuous variable, which is not realistic in physical devices.  To address this issue, we employ a permittivity capping scheme during optimization.  We define a maximum permittivity `$\epsilon_m$' corresponding to a material of interest. During the iterative process, if the relative permittivity of any cell becomes either less than $1$ (vacuum) or greater than $\epsilon_m$, that cell is pushed back into the acceptable range.  It was found that with this capping scheme, the structures converged to binary (each pixel being either vacuum or material with a permittivity of $\epsilon_m$) after a number of iterations without specifying this choice of binary materials as a requirement of the optimization.  Therefore, only minimal post-processing of the structures was required.

\begin{figure}[htb!]
\centering\includegraphics[width=\textwidth]{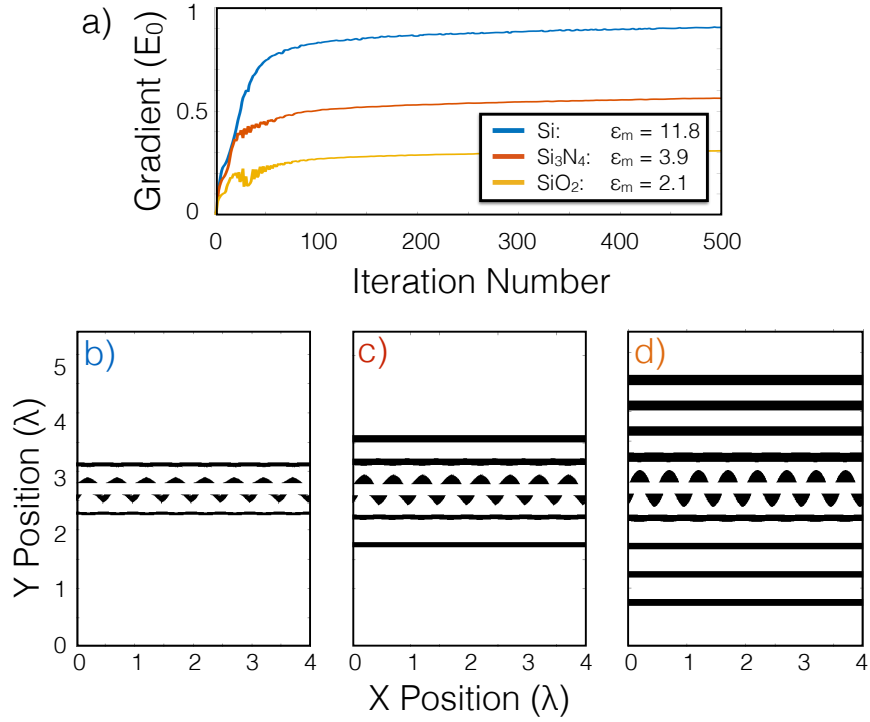}
\caption{Demonstration of the structure optimization for $\beta=0.5$, laser wavelength $\lambda=2\,\mu$m, and a gap size of $400$\,nm. A plane wave is incident from the bottom in all cases.  (a) Acceleration gradient as a function of iteration number for different maximum relative permittivity values, corresponding to those of Si, Si$_3$N$_4$, and SiO$_2$ at the laser wavelength.  The optimizations converge after about five-hundred iterations.  (b-d) Final structure permittivity distributions (white = vacuum, black = $\epsilon_m$) corresponding to the three curves in (a).  Eight periods are shown, corresponding to four laser wavelengths.  For each (b-d), design region widths on each side of the particle gap were given by $1$, $2$, and $4\,\mu$m for Si, Si$_3$N$_4$, and SiO$_2$, respectively.}
\label{fig:optimization-results_beta0.5}
\end{figure}

The results of this optimization scheme are shown in Fig. \ref{fig:optimization-results_beta0.5}(b-d) for three different $\epsilon_m$ values corresponding to commonly explored DLA materials.  The design region was taken to be a rectangle fully surrounding but not including the particle gap.  The design region was made smaller for higher index materials, since making it too large led to divergence during the iteration.  We found that a totally vacuum initial structure worked well for these optimizations.  However, initially random values between 1 and $\epsilon_m$ for each pixel within the design region also gave reasonable results.  

This optimization scheme seems to favor geometries consisting of a staggered array of field-reversing pillars surrounding the vacuum gap, which is already a popular geometry for DLA. However, these optimal designs also include reflective mirrors on either side of the pillar array, which suggests that for strictly higher acceleration gradients, it is useful to use dielectric mirrors to resonantly enhance the fields in the gap.

It was observed that for random initial starting permittivity distributions, the same structures as shown in Fig. \ref{fig:optimization-results_beta0.5} are generated every time.  Furthermore, these geometries are remarkably similar to those recently proposed \cite{ACHIP_meeting3}, although these designs do not include the reflective front mirror.  These findings together suggest that the proposed structures may be close to the globally optimal structure for maximizing $G$.

It was further found that convergence could be achieved faster by a factor of about ten by including a `momentum' term in the update equation.  This term corresponds to the sensitivity calculated at the last iteration multiplied by a constant, $\alpha' < 1$. Explicitly, for iteration number `$j$' and pixel `$i$' 

\begin{equation}
\epsilon_i^{ (j+1) } := \epsilon_i^{ (j) } + \alpha\Bigg[ \frac{dG}{d\epsilon_i}^{ (j) } + \alpha'\frac{dG}{d\epsilon_i}^{ (j-1) }\Bigg].
\end{equation}

\subsection{Acceleration Factor Maximization}

DLAs are often driven with the highest input field possible before damage occurs.  Therefore, another highly relevant quantity to maximize is the `acceleration factor' given by the acceleration gradient divided by the maximum electric field amplitude in the structure.  This quantity will ultimately limit the amount of acceleration gradient we can achieve when running at damage threshold.  Explicitly, the acceleration factor is given by

\begin{equation}
f_{A} = \frac{G}{\max\{|\vec{E}|\}}.
\end{equation}
Here, $|\vec{E}|$ is a vector of electric field magnitudes in our structure, and the $\max\{\}$ function is designed to pick out the highest value of this vector in either our optimization or material region, depending on the context.  We would like to use the same basic formalism to maximize $f_{A}$.  However, since the $\max\{\}$ function is not differentiable, this is not possible directly.  Instead we may use a `smooth-max' function to approximate $\max\{\}$ as a weighted sum of vector components

\begin{equation}
\max\{|\vec{E}|\} \approx \frac{\sum_i |\vec{E}_i| \ \textrm{exp}\big({a|\vec{E}_i|}\big)} {\sum_i \textrm{exp}\big({a|\vec{E}_i|}\big)}.
\end{equation}
Here, the parameter $a \geq 0$ controls the relative strength of the exponential sum terms, for $a = 0$, this function simply gives the average value of the field amplitudes.  By sweeping $a$ and examining the acceleration factors of the resulting optimized structures, we determined that $a = 3$ gave the best improvement in $f_A$.  If $a$ is too large, the calculation may induce floating point overflow or rounding error issues.  

\begin{figure}[htb!]
\centering\includegraphics[width=\textwidth]{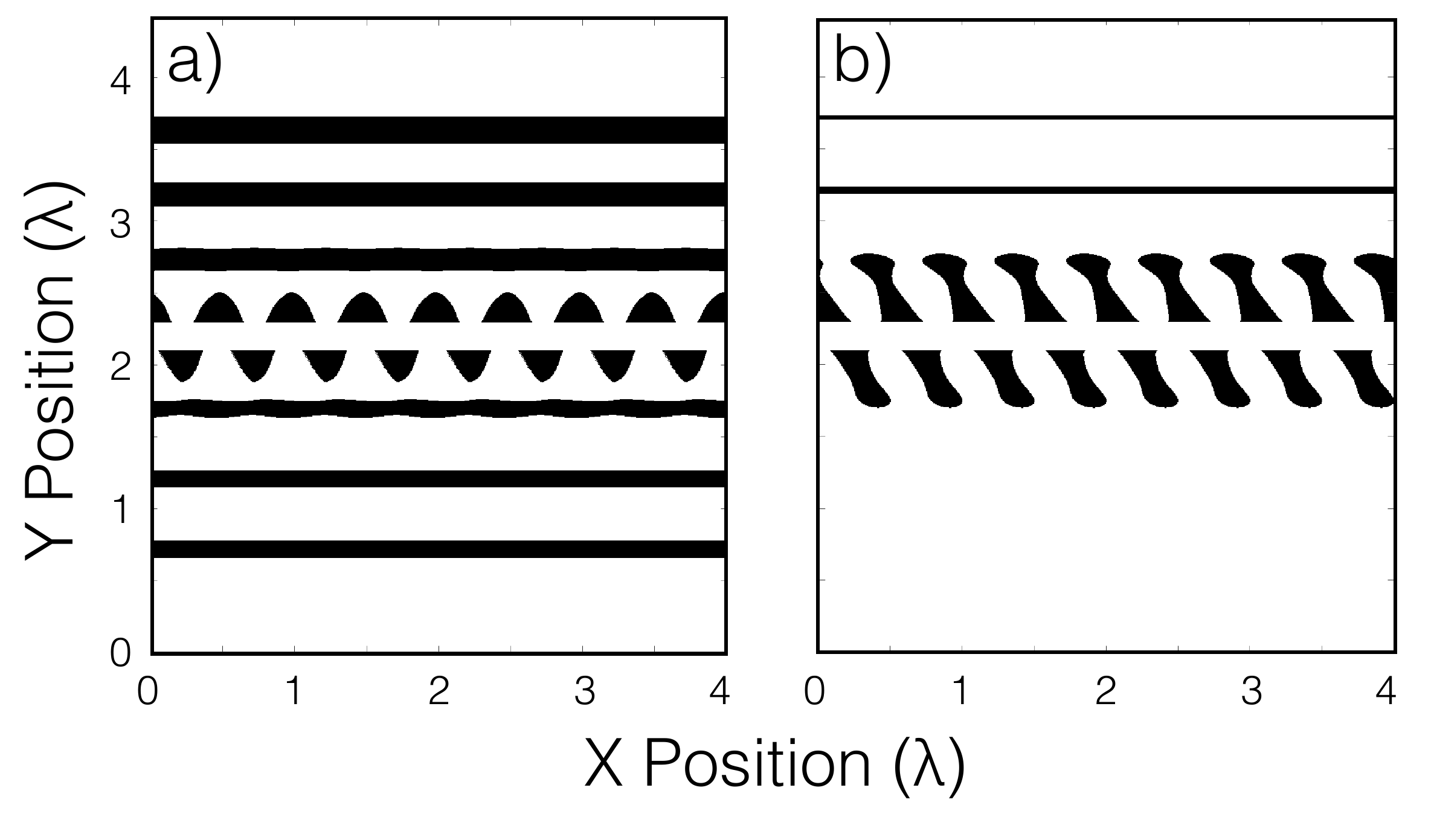}
\caption{Demonstration of the final structures after optimization for (a) maximizing gradient only, (b) maximizing the acceleration factor.  $\beta = 0.5$, laser wavelength $\lambda=2\,\mu$m, gap size of $400$\,nm.  $\epsilon_m = 2.1$, corresponding to SiO$_2$.  In (a), the high gradients are achieved using reflective dielectric mirrors to confine and enhance the fields in the center region.  In (b), these dielectric mirrors are removed and the pillar structures are augmented.  The structure in (b) shows a 23\% increase in the acceleration factor in the material region when compared to (a).}
\label{fig:optimization-results_fAC}
\end{figure}

Using this smooth-max function, one may calculate $\frac{df_{A}}{d\epsilon_i}$ analytically and perform structure optimizations in the same way that was discussed previously.  The derivation of the adjoint source term is especially complicated and omitted for brevity, although the end result is expressed solely in terms of the original fields, the adjoint fields, and the $\frac{d\hat{A}}{d\gamma}$ operator, as before.  Two structures with identical parameters but optimized, respectively, for maximum $G$ and $f_A$ are shown in Fig. \ref{fig:optimization-results_fAC}.  On the left, we see that the $G$ maximized structure shows the characteristic dielectric mirrors, giving resonant field enhancement.  On the right is the structure optimized for $f_A$, which has eliminated most of its dielectric mirrors and also introduces interesting pillar shapes.  In Table \ref{tb:table} the main DLA performance quantities of interest are compared between these two structures.  Whereas the acceleration gradient is greatly reduced when maximizing for $f_A$, the $f_A$ value itself is improved by about 25\% or 23\% depending on whether one measures the maximum field in the design region or the material-only region, respectively.  These findings suggest that the AVM strategy is effective in designing not only resonant, high acceleration gradient structures, but also non-resonant structures that are more damage resistant.  In the future, when more components of DLA are moved on-chip (such as the optical power delivery), it will be important to have control over the resonance characteristics of the DLA structures to prevent damage breakdown at the input facet.  Our technique may be invaluable in designing structures with tailor-made quality factors for this application.

\begin{table}[ht!]
\centering
\caption{Acceleration factor ($f_A$) before and after maximization.}
\begin{tabular}{lccc}
\hline
Quantity & Value (max $G$) & Value (max $f_A$) & Change  \\ \hline
Gradient ($E_0$) & 0.1774 & 0.0970 & -45.32\%\\ 
$\max\{|\vec{E}|\}$ in design region & 4.1263 & 1.7940 & -56.52\%\\ 
$\max\{|\vec{E}|\}$ in material region & 2.7923 & 1.2385 & -55.84\% \\ 
$f_A$ in design region & 0.0430 & 0.0541 & +25.81\% \\
$f_A$ in material region & 0.0635 & 0.0783 & +23.31\% \\ \hline
\end{tabular}
\label{tb:table}
\end{table}

\section{Discussion}
%1. Major findings
We found that AVM is a reliable method for optimizing DLA structures for both maximum acceleration gradient and also acceleration factor.  The optimization algorithm discussed shows good convergence and rarely requires further post-processing of structures to create binary permittivity distributions.  Therefore, it is a simple and effective method for designing DLAs.
%2. What is new and novel
Whereas most structure optimization in this field uses parameter sweeps to search the design space, the efficiency of our method allows us to more intelligently find optimal geometries without shape parameterization.  Furthermore, the structures that we design are fabricable.  
%3. How does it compare with literature?
Although no DLA structures have been tested at the proposed wavelength of $2\, \mu$m, both simulations \cite{plettner2006proposed} and experimental results from other wavelengths \cite{leedle2015dielectric} show gradients far below those presented here.
%4. What are the limitations?
We had limited success designing DLA structures in the relativistic ($\beta \approx 1$) regime, especially for higher index materials, such as Si.  We believe this is largely due to the stronger coupling between electron beam and incident plane wave at this energy.  The characteristics of the adjoint source change dramatically at the $\beta = 1$ point.  Whereas in the sub-relativistic regime, the adjoint source generates an evanescent near-field extending from the gap particle position, at $\beta \geq 1$, the adjoint fields become propagating by process of Cherenkov radiation. Upon using the above described algorithm, the gradients diverge before returning to low values, no matter the step size $\alpha$.  The only way to mitigate this problem is to choose prohibitively small design regions or low index materials, such as SiO$_2$.  

The AVM formalism presented in this work may also be extended to calculate higher order derivatives of $G$.  For each higher order, the form of the derivative of $G$ can be derived in a fashion very similar to the one outlined for first order.  Given the full Hessian $H_{i,j} = \frac{d^2G}{d\epsilon_i d\epsilon_j}$, as calculated by AVM, one could use Newton's method to do optimization.  However, to perform exactly, this calculation would require as many additional simulations as there are grid points within the design region.  Therefore, these higher order methods are inconvenient for our purposes where there are generally tens of thousands of design pixels.  This limitation may be averted by using approximate methods for finding the inverse Hessian \cite{nocedal1980updating}, which may provide substantial improvement to optimization results and convergence speeds in certain cases.  However, in our case there was no need to explore beyond first order due to the relative success and speed of the algorithm presented.

%5. Implications/future work
As future works, our goal is to fabricate and test these structures experimentally, as well as include further metrics into the optimization if necessary, such as favoring larger feature sizes and incorporating focusing effects.  Furthermore, this method is of great interest in designing waveguide-coupled accelerator structures, where typical designs optimized for plane wave input are likely suboptimal.  This will be of critical importance when moving the optical power delivery source on-chip.  

In addition to the side-incident geometry explored, this technique is applicable to designing other dielectric-based accelerator structures.  This includes particle-laser co-propagating schemes \cite{cowan2008three} and perhaps dielectric wakefield acceleration  \cite{zhang1997stimulated}, among others.  Therefore, we expect that our results may find use in the larger advanced accelerator community.

\section{Conclusion}

We have introduced the adjoint variable method as a powerful tool for designing dielectric laser accelerators for high gradient acceleration and high acceleration factor.  We have further shown that the adjoint simulation is sourced by a point charge flowing through the accelerator, which quantifies the reciprocal relationship between an accelerator and a radiator.

Optimization algorithms built on this approach allow us to search a substantially larger design space and generate structures that give gradients far above those normally used for each material.  Furthermore, the structures designed by AVM are fundamentally not constrained by shape parameterization, allowing never-before-seen geometries to be generated and tested.

\section*{Funding}
%ACHIP
This work was supported by the Gordon and Betty Moore Foundation under grant GBMF4744 (Accelerator on a Chip), %SLAC
and by the U.S. Department of Energy under Contract DE-AC02-76SF00515.

\section*{Acknowledgments}
The authors thank Zhexin Zhao, Yu (Jerry) Shi, and Wonseok Shin for illuminating discussions.  We also thank members of the ACHIP collaboration for feedback and support in developing this work.

\end{document}